\documentclass[final,1p,times]{elsarticle}

\usepackage{amssymb}
\PassOptionsToPackage{hyphens}{url}\usepackage{hyperref}
\usepackage{xcolor}
\usepackage{caption}
\usepackage{enumitem}
\usepackage{float}
\usepackage{subcaption}
\hypersetup{
	colorlinks=true, 
	linktocpage=true,
	breaklinks=true, 
	pageanchor=true,
	plainpages=false, 
	hypertexnames=true,
	urlcolor=blue, 
	linkcolor=blue,
	citecolor=blue, 
	pdftitle={Assessing place-based activity response to COVID-19 policy: A country comparison},
	pdfauthor={Grant McKenzie, Benjamin Adams},
	pdfsubject={Assessing place-based activity response to COVID-19 policy: A country comparison},
	pdfkeywords={COVID-19, mobility, policy, community response, activity pattern},
	pdfcreator={pdfLaTeX},
	pdfproducer={LaTeX via TeX Live on GNU/LINUX}
}

\journal{arXiv.org}

\begin{document}

\begin{frontmatter}



\title{A country comparison of place-based activity response to COVID-19 policies}


 \author[label1]{Grant McKenzie}
 \address[label1]{McGill University, Montr\'eal, Canada}

\author[label2]{Benjamin Adams}
\address[label2]{University of Canterbury, Christchurch, New Zealand}

\begin{abstract}

The emergence of the novel Coronavirus Disease in late 2019 (COVID-19) and subsequent pandemic led to an immense disruption in the daily lives of almost everyone on the planet.  Faced with the consequences of inaction, most national governments responded with policies that restricted the activities conducted by their inhabitants.  As schools and businesses shuttered, the mobility of these people decreased.  This reduction in mobility, and related activities, was recorded through ubiquitous location-enabled personal mobile devices.  Patterns emerged that varied by place-based activity.  In this work the differences in these place-based activity patterns are investigated across nations, specifically focusing on the relationship between government enacted policies and changes in community activity patterns.  We show that people's activity response to government action varies widely both across nations as well as regionally within them.  Three assessment measures are devised and the results correlate with a number of global indices.  We discuss these findings and the relationship between government action and residents' response.

\end{abstract}

\begin{keyword}
	COVID-19\sep mobility \sep policy \sep community response \sep activity pattern


\end{keyword}

\end{frontmatter}


\section{Introduction}
\label{sec:introduction}

The emergence and global spread of the novel Coronavirus Disease (COVID-19), caused by the SARS-CoV-2 virus, will have a lasting impact on human life. 
In late 2019, the disease was identified in Wuhan, China and quickly spread across the globe~\cite{huang2020clinical}.  Nations responded in a variety of ways as cases of the disease increased dramatically within their populations.  Some governments responded immediately, severely restricting residents' mobility and pausing most economic activity~\cite{kupferschmidt2020can} while others 
chose to implement less restrictive policies~\cite{sjodin2020covid}.  Across the globe, borders began to close and many national governments chose to close schools and businesses, essentially shutting down local economies, while others limited their response to simple physical distancing practices~\cite{glass2006targeted}.  The speed at which countries responded to this pandemic also varied drastically with some governments taking action within days of their first confirmed cases, and others waiting to implement public health policies until a pattern could be identified.  Still other 
governments have yet to instrument meaningful COVID-19-related policy choosing instead to make an argument for ``herd immunity''~\cite{kwok2020herd}.

At the heart of the pandemic, however, are the inhabitants of these different countries.  People look to their government for information on the severity and spread of the disease and rely on what they are told to make informed decisions on how to act. For most, government enacted policies dictate how they will conduct their daily affairs 
during a global health emergency.  As officials around the world began to cancel public events and limit public transit options, inhabitants were forced to respond by limiting their mobility and changing their activity behavior. A decade ago, this change in community mobility would have been measured at a large scale through survey samples of the population.  Today, however, this change in activity behavior is being monitored and recorded in real-time through context-aware technology.  The ubiquity of location-enabled mobile devices means that mobile device manufacturers (e.g., Apple, Google) and data providers (e.g., Verizon, Deutsche Telekom) have unlocked access to the mobility activity of people around the planet to a degree previously unattainable. 

During this crisis, a few of these companies have offered limited public access to anonymized aggregations of this data.  Apple, for example, is reporting a decrease in human mobility globally based on route searches performed by users of their \textit{Maps} platform over the past few month~\cite{applemaps}.  In some instances route queries have dropped below 80\% of their baseline suggesting that a significant amount of users have reduced their mobility due to the pandemic.  Similar results have been released by the Descartes Lab, an analytics company, for counties in the United States~\cite{warren2020mobility}.  Aside from purely spatial movement data, other companies have focused on discrete places, using location-enabled mobile device data to identify the types of places (e.g., Bank, Grocery Store, Restaurant) people visit, and how their visiting behavior has changed over the past few months.  For instance, Foursquare~\cite{foursquare} has demonstrated the impact of COVID-19 on place-based social media check-ins within the United States\footnote{https://visitdata.org/index.html} and SafeGraph~\cite{safegraph} is publishing similar data based on a panel of location-based mobile surveys.

Aside from the Apple Maps data, however, the vast majority of data published on changes in human activity behavior has focused exclusively on the United States.  In this work, we are interested in how the residents of different countries respond to their government's policies related to COVID-19, therefore identifying the place-based activity responses of people from a range of countries is essential.  With this goal in mind, we make use of the \textit{Google Community Mobility Dataset}.  This dataset is an aggregate of the place-based activity behavior of millions of individuals as collected through their location-enabled mobile devices.  If a mobile device user has \textit{Google Location Services} enabled (required when using the Google Maps application), their location data is anonymously collected, aggregated with other users across a region (e.g., Country or District), and reported in one of six place type categories. Further information on these categories is discussed in Section~\ref{sec:data1}.  Not only are these data based on one of the most widely used location-based services on the planet, but the data is also passively collected thus reducing the bias of the mobility patterns, as compared to active collection such as surveys or ``check-ins''~\cite{mckenzieuncovering}. For the purposes of this research, this data is an ideal representation of activity response to the COVID-19 pandemic and forms the foundation on which we conduct our analysis.   



As of early May 2020 (time of writing) most governments have responded to the crisis to the degree they are likely to, and we are currently observing many governments announce plans to lift lock-downs, relax mobility restrictions, and re-open their economies.  Now is an opportune time to examine the relationship between government action and resident response with respect to place-based activities.  Understanding how residents of certain countries responded to their government in a time of crisis is essential for predicting how people may react in future crises and exposing the dynamic between politicians and their constituents in difficult times. It also helps us to identify which countries respond to crises in similar (or different) ways.  With these objectives in mind, our analysis focuses on developing three \textit{assessment measures} on which to compare and contrast countries through their response to the COVID-19 pandemic.  These assessment measures will be developed by addressing the following five research questions (\texttt{RQ}).


\begin{enumerate}[label=\texttt{RQ\arabic*}]

	\item \label{rq1} Is there a quantifiable relationship between the policies enacted by a country's government and the place-based activity response from their inhabitants?  For instance, is the magnitude of policy action mirrored by the magnitude of activity response? Do countries differ in their responses? This question helps us understand whether the changes in activity patterns that we see world-wide are actually due to the imposition of government policies, or perhaps due to some other reason, e.g., international media. The differences between countries give insight into differences among governments in their ability to enact the envisioned policies.
	
	\item \label{rq2} Which of the six place-based activity categories, as reported by Google, are most affected by policy action?  Do these remain consistent between countries? This helps us to identify which policies are most effective for influencing human behavior in order to limit the spread of COVID-19, and also which policies are less effective. Since limiting different kinds of activities will lead to different economic outcomes, better understanding this relationship helps us optimize our responses for the desired outcome.
	
	\item \label{rq3} Is there a measurable temporal lag between policy enactment and activity response?  If so, does this lag vary between countries? During an exponential growth phase of a disease such as COVID-19 a delay of just a few days can make a large difference in the outcome. Understanding temporal lags between policy enactment and activity response is critical to inform policy-makers who are trying to quickly enact social distancing policies.
	
	\item \label{rq4} Does an increase in activity pattern variability within a country (at the subnational level) correlate with a decrease in similarity to government response (\ref{rq1})? This helps us to understand whether a unified, national response is more or less effective than one that is managed by regional and local governments.
	
	\item \label{rq5} How does the relationship between government policy action and place-based activity response correlate with global indices (such as the Development Index or Corruption Perception Index)? This last research question gives some insight into whether there are distal causes based on the socio-economic and political conditions within countries that influence how place-based activity manifests under government imposed lockdown conditions.
	
\end{enumerate}

The remainder of this paper is organized as follows. In Section~\ref{sec:relatedwork} we discuss existing work on this topic and related topics.  An overview of the data used in this analysis is presented in Section~\ref{sec:data}.  Section~\ref{sec:response} presents the methodology used in developing the assessment measures and showcases how these measures can be applied to compare and contrast nations.  This is followed by Section~\ref{sec:indices} where we compare our three assessment measures with existing global indices.  Finally, our findings are discussed in Section~\ref{sec:discussion} and future work and conclusions are presented to the reader.
\section{Related Work}
\label{sec:relatedwork}

We are still in the early days of the COVID-19 pandemic.  A burgeoning domain of biomedical research has developed around the virus itself with many researchers just now beginning to study the global societal impacts of the disease.  Much of the emerging peer-reviewed work has focused on the regions with early outbreaks.  One such study by Kraemer et al.~\cite{kraemer2020effect} studied the correlation between real-time mobility data from the internet service company Baidu and the spread of the disease in Wuhan, China.  The authors found that once the national government established control measures, the correlation decreased substantially.  Further work on the transmission rate of the disease found that travel restrictions enacted by the government delayed the epidemic progression by 3-5 days within China, but significantly slowed the spread elsewhere in the world~\cite{chinazzi2020effect} pointing to the global impacts of the policy actions taken by a single country.

Much of the \textit{response} research being conducted on COVID-19 is still in its infancy, but we can look to other recent global events to understand how nations and their people respond.  There is an existing body of literature exploring the impact of natural disasters~\cite{wang2014quantifying}, terrorism~\cite{nail2016tale}, and economic crises~\cite{ishikawa2011impact} on human mobility and activity patterns.  Most of these studies define human mobility at the scale of migration between localized regions and little research has explored the day-to-day impact on the mobility of individuals or groups at a multi-national level, brought about by a global crisis.

In attempting to find parallels to the current pandemic, we turn to research on the response to previous global pandemics.  There have been a series of studies that explore the impact of human movement on the spread of diseases including influenza~\cite{merler2010role}, ebola~\cite{wesolowski2014commentary,halloran2014ebola}, and infectious diseases in general~\cite{kraemer2019utilizing}.  Many of these studies make use of local population datasets including mobile device call detail records (CDRs), location histories, and the results of travel/mobility surveys.  For instance, Dallatomasina et al.~\cite{dallatomasina2015ebola} used cellphone records to track the transmission of ebola in rural West Africa with some success.  Early research out of Brazil has shown this methodology to be useful in tracking the spread of COVID-19~\cite{queiroz2020large}.  Much of this work is a pre-cursor to mobile device-based contact tracing applications that are currently in the works~\cite{ferretti2020quantifying}.  Few studies, however, have had access to the breadth or volume of data now being collected and published by private technology companies as a response to this specific pandemic. 

As companies release their data to researchers, a growing body of work is emerging related to COVID-19 mobility response.  The vast majority of this work is focused in the United States with very limited work concentrated at the global scale.  Visualization tools for supporting analytics have been a target for many researchers~\cite{gao2020mapping,dong2020interactive}.  Other efforts have investigated the relationship between political views and compliance with government policy~\cite{painter2020political,merkley2020rare}.  U.S.-focused research has also demonstrated that government policy that limits every day life has been shown to be having an impact on many sectors of the economy~\cite{atkeson2020will}.  A recent study of high school students suggested that school closures and concerns over COVID-19 are significantly impacting performance of students in their subject areas~\cite{viner2020school} and will continue to have a lasting impact on our education system.  While limited data is available, similar educational and economic impacts are also being found outside of the United States~\cite{sintema2020effect}.

Similar work to what we present here has investigated certain economic factors that contribute to a change in place-based activity patterns (also using Google community mobility patterns)~\cite{m2020democracy,morita2020international}.  These efforts, however, did not look at government policy as a whole or the correlation of response over time.  Other researchers have used this data to predict specific country-level responses in the future~\cite{russo2020tracing,morita2020changes}. To the best of our knowledge, no research has investigated the relationship between government policy action and place-based activity response at the scale we present here, or using the range of methods we propose.

\section{Data}
\label{sec:data}

The data used in these analyses are publicly available.  For reproducibility, links to data sources have been provided as footnotes where appropriate.  Data produced from our analysis are available at \url{https://platial.science/covid19code}. 

\subsection{Place-based Activity Patterns}
\label{sec:data1}

\begin{figure}[h]
	\centering
		\includegraphics[width=0.7\textwidth]{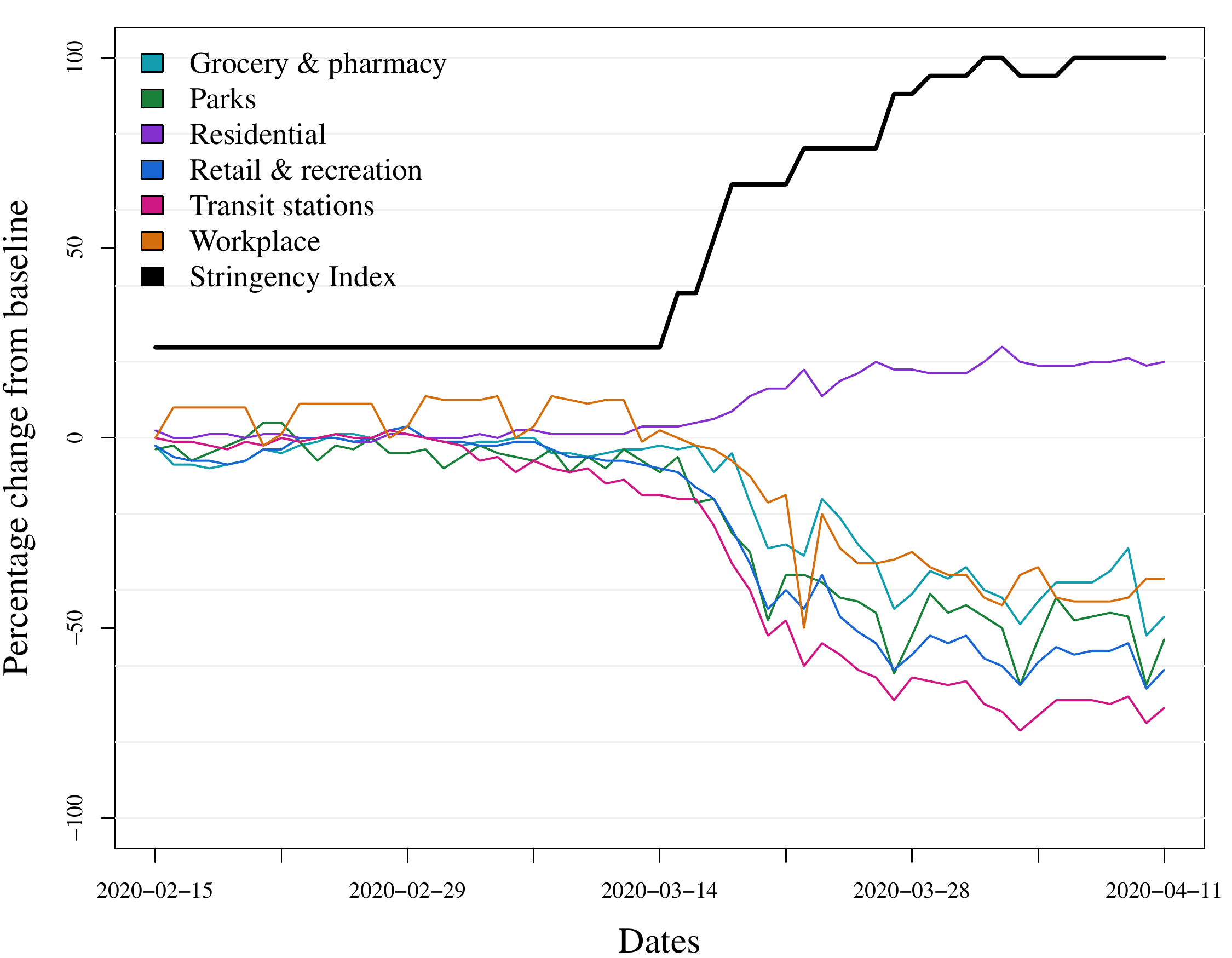}  
	\caption{An example (Oman) of a change in place-based activity patterns over time as reported through Google's location services.  The stringency index is also shown in this example as a thick black line.}
	\label{fig:example}
\end{figure}

In early April 2020 Google began publishing static reports showing daily change in activity patterns starting from February 15th~\cite{googledata}.  These reports, which Google call their \textit{Community Mobility Reports},\footnote{\url{https://www.google.com/covid19/mobility/}} present plots containing daily percentage change from baseline (baseline being January 3--February 6, 2020) for six activity categories in 129 countries.  The place types that contribute to these activity categories were determined by Google and labeled as \textit{Grocery \& pharmacy}, \textit{Parks}, \textit{Transit stations}, \textit{Retail \& recreation}, \textit{Residential}, \textit{Workplaces}.  Brief descriptions of these categories are available in the Google community mobility documentation.\footnote{\url{https://www.google.com/covid19/mobility/data_documentation.html}}  In late April, Google began providing the raw values for these data along side the original PDF reports.  For this study, we are using activity reports that cover an eight week time period from February 15 to April 11, 2020.  The maximum baseline change in the negative direction was 100\% and 497\% in the positive direction (an outlier due to a country-wide festival). An example of these activity patterns are shown in Figure~\ref{fig:example} along with the stringency index for the country (introduced in the next section). 

\subsection{Government Policy and Stringency Index} 

As nations around the world were faced with the spreading COVID-19 pandemic, their governments responded with a range of measures and policy decisions.  A team of researchers at the Blavatnik School of Government, University of Oxford developed a tool to track these measures and produce a response dataset\footnote{\url{https://www.bsg.ox.ac.uk/research/research-projects/coronavirus-government-response-tracker}} that allows national policy responses to be compared across a range of measures.  The key product of this \textit{Coronavirus Government Response Tracker} is the \textit{Stringency Index}.  This index is made of up 13 indicators of government response, recorded daily for 149 countries.  Nine of these indicators assign an ordinal value to policy action related to school closures, travel bans, public event cancellations, etc.  The other three focus on financial or monetary measures.  The \textit{Stringency Index} is calculated through a weighted combination of these indicators with a minimum value of 0 and maximum of 100.   The index was calculated back to January 1, 2020 and through to at least April 11 (the end of our temporal analysis window).  The full methodology for how this stringency index was developed is available at~\cite{oxford2020}.  The dataset is currently being used in a range of preliminary COVID-19 related analysis~\cite{gustafsson2020does,barberia2020confronting,elgin2020economic}.
 

\subsection{Indices and Country Information}

Finally, we compare the results of our analysis to a number of other country-specific attributes and global indices.  The latest population counts, areas, and population densities for each country were accessed from GeoNames.org.\footnote{\url{https://www.geonames.org/countries/}} The number of confirmed cases and deaths due to COVID-19 were downloaded from the World Health Organization Coronavirus dashboard.\footnote{\url{https://covid19.who.int/}}  Two indices of global development were used, the United Nations Human Development Index~\cite{hdr2019} and the World Bank World Development Indicators.\footnote{\url{https://databank.worldbank.org}} Finally, the results of our analysis were compared against the Corruption Perception Index~\cite{transparencyInternational} and the World Press Freedom Index published by Reporters Without Borders.\footnote{\url{https://rsf.org/en/ranking}}




\section{Assessing place-based activity response to government policy action}
\label{sec:response}

In this section we discuss the methodology used to assess the relationship between the stringency index, representing a country's government policy action, and the place-based activity patterns recorded via mobile devices, representing inhabitant response.  We establish three methods of assessing the response: similarity in magnitude of response, lag response time, and subregional variability of the response.

\subsection{Similarity}

Two different methods, \textit{cosine similarity} and \textit{Pearson correlation}, were initially used in addressing \ref{rq1}---assessment of the similarities between the stringency index and activity patterns.  The cosine similarity approach views both sets of data as a set of vectors, measuring the cosine of the angle between each pair and producing a similarity value bounded between 0 and 1. While both cosine similarity and Pearson's correlation are fundamentally variations on the inner product, they vary based on centering.  For our purposes, cosine similarity is the most appropriate method for comparing activity patterns to government policy over time due to the fact that it is not invariant to shifts, a feature that is not true for Pearson's correlation. We will use both of these approaches, however, for assessing similarity and report later on the correlation.

Given our hypothesis that activity patterns decrease over time in response to an increase in the stringency index, we first inverted five of the six mobility patterns, namely \textit{Retail \& recreation}, \textit{Grocery \& pharmacy}, \textit{Workplace}, \textit{Transit stations}, and \textit{Parks} in order to conduct \textit{similarity} assessment (as opposed to dissimilarity).  These five activity patterns are expected to decrease with an increase in government policy.  The \textit{Residential} pattern, on the other hand, is expected to increase as government restrictions increase, so this was not inverted. 

\subsubsection{Assessing activity patterns across countries}

After flipping five of the activity patterns, cosine similarity and Pearson's correlation were calculated between each of the six categories and the stringency index for each country.  Averaging across all countries, the cosine similarities are shown in Table~\ref{tab:cosagg}.

\begin{table}[h]
	\centering
	\small
	\begin{tabular}{lrrr}
		\hline
		\textbf{Activity Category}    & \textbf{Mean} & \textbf{Median} & \textbf{Standard Deviation} \\
		\hline
Transit stations   & 0.945         & 0.959           & 0.048                       \\
Residential       & 0.937         & 0.949           & 0.045                       \\
Retail \& recreation  & 0.936         & 0.947           & 0.051                       \\
Workplace         & 0.859         & 0.909           & 0.179                       \\
Grocery \& pharmacy   & 0.741         & 0.807           & 0.234                       \\
Parks             & 0.707         & 0.897           & 0.450                          
\\
		\hline          
	\end{tabular}
	\caption{Average cosine similarity comparing each place-based activity category to the stringency index across all countries.}
	\label{tab:cosagg}
\end{table}

The results of this analysis identify differences in the relationship of each activity pattern to the stringency index (\ref{rq2}).  Three categories, \textit{Transit stations}, \textit{Residential}, and \textit{Retail \& recreation}, all report relatively high similarity and low standard deviations suggesting that these similarities are relatively consistent across all countries.  The \textit{Workplace} activity category clearly aligns with the stringency index though not to the same degree as the previously mentioned three.  Some potential reasons for this are explored in Section~\ref{sec:discussion}. As expected, \textit{Parks} are least similar to the stringency index, and with a high standard deviation, which indicates there is considerable variation in this mobility pattern across countries.  The \textit{Grocery \& pharmacy} category is not as close to the stringency index with a larger standard deviation across countries than one might expect.  Again, some potential reasons for this are discussed in Section~\ref{sec:discussion}.  Notably, the ranked order similarity for activity categories based on averaged Pearson correlation is the same as that of cosine similarity.

\subsubsection{Assessing countries across activity patterns}

The previous section focused on the six categories and averaged across countries, here we average across activity patterns and compare individual country responses.  We calculated the mean of the cosine similarity between each activity pattern and the stringency index for all activity categories excluding \textit{Parks} (due to the volatility and weather dependency of this category). This resulted in a single similarity value for each country.  We also calculated the average Pearson's correlation coefficient value in a similar manner.  Provided a cosine similarity and Pearson's correlation for each country, we computed Kendall's $\tau$ to measure concordance between the two measures.  This produced a $\tau$ correlation coefficient of 0.602 (p $<$ 0.01), indicating a high level of concordance between these two approaches.  

\begin{figure}[h]
	\centering
	\includegraphics[width=1\textwidth]{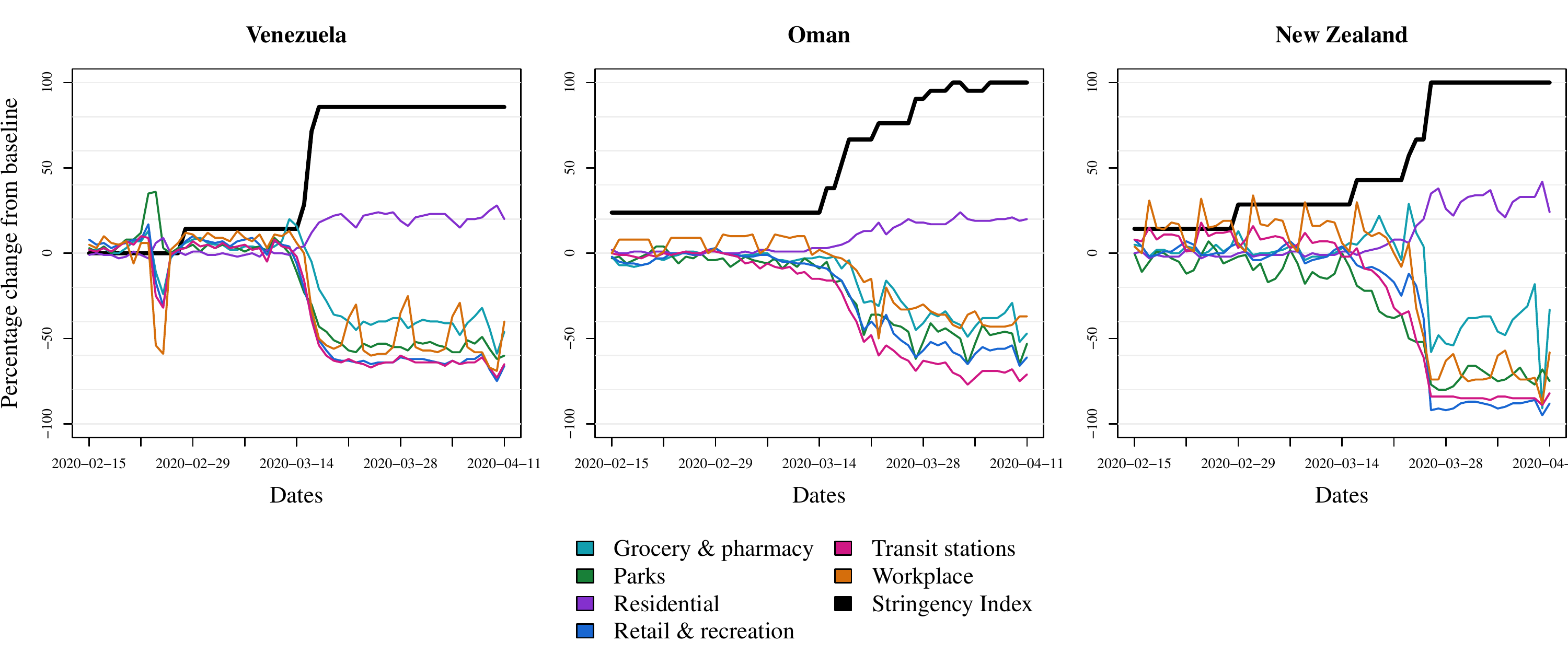}  
	\caption{Highest average cosine similarity between stringency index and five place-based activity patterns.}
	\label{fig:topthree}
\end{figure}

\begin{figure}[h]
	\centering
	\includegraphics[width=1\textwidth]{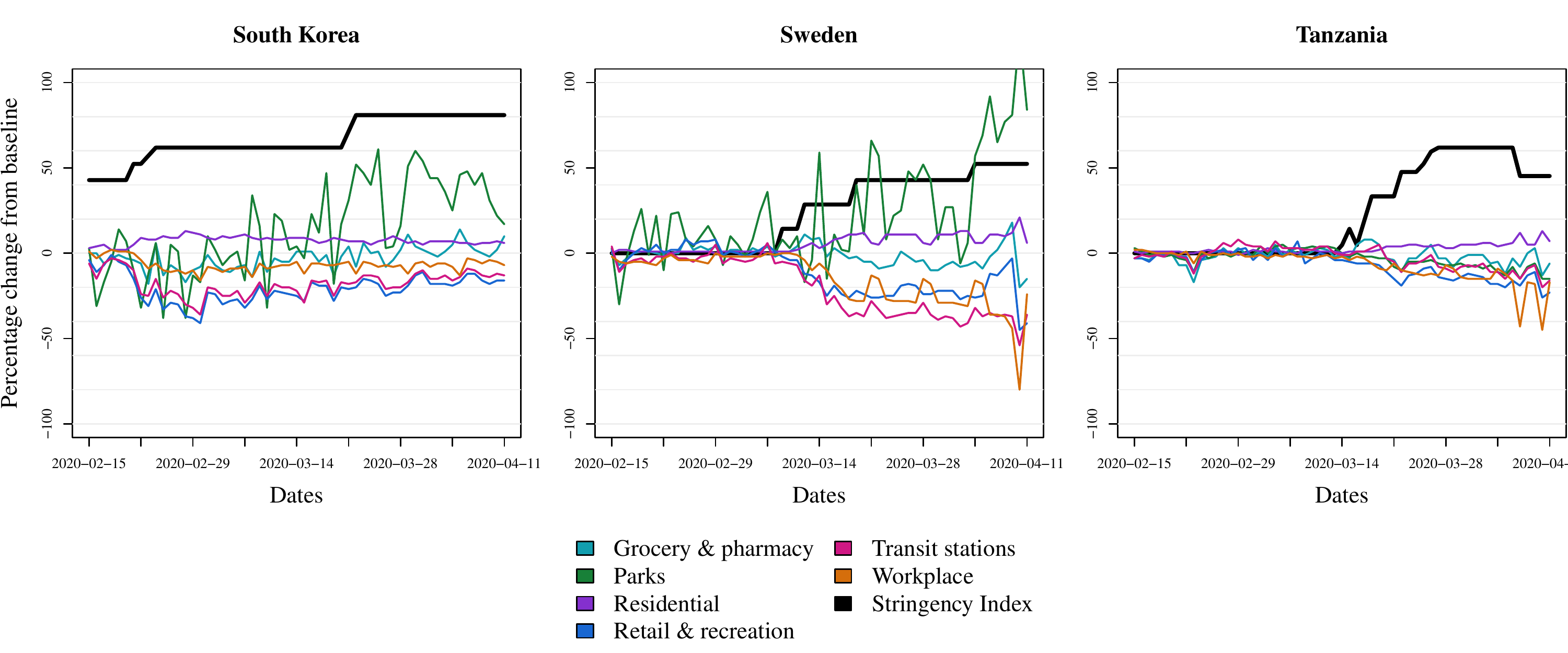}  
	\caption{Lowest average cosine similarity between stringency index and five place-based activity patterns.}
	\label{fig:bottomthree}
\end{figure}

Countries were then ranked based on their average activity pattern to stringency index similarity.  In essence, this similarity value provides a relative indication of how a population responds to COVID-19-related policy enacted by its government. A high average similarity suggests that an increase in government policy leads to a decrease in overall place-based activity.  Plots for three of the countries with the highest average similarity are shown in Figure~\ref{fig:topthree} with three of the least similar shown in Figure~\ref{fig:bottomthree}.  The full rank of countries is available in the project data directory at \url{https://platial.science/covid19code}. 

Visually we see a high degree of similarity in the magnitude and absolute slope of the lines in these Figures, representing the actions taken by the government and the response from residents.  Exploring the countries with the least similar patterns, the country with the lowest value was South Korea.  Further investigation found that this is solely due to the time window of our analysis.  The peak of new cases in South Korea occurred on February 29, 2020 and local health officials started testing airport staff on January 21, 2020.  This is supported by the high stringency index value of 60 as it enters our analysis window on February 15th.  Similarly, Japan, Taiwan, Mongolia, and Hong Kong all show high stringency indices entering our analysis window suggesting that if a substantial decrease in residents' place-based activity occurred, it likely took place prior to February 15th and adjustments from the baseline (the previous months as mentioned in the Data section) would be quite small.  The first country that entered our time analysis window with a relatively stable stringency index near zero was Sweden followed closely by all other Scandinavian countries. This supports reporting by journalists and policy risk analysts that Scandinavian governments, specifically Sweden, are approaching the pandemic differently than most other countries~\cite{nyt2020,wt2020}.  Tanzania (Figure~\ref{fig:bottomthree}c) is the country with the lowest cosine similarity between citizen activity response and government stringency index outside of Asia or Europe.

\subsection{Lag response time}

The previous assessment method reported the similarity between the stringency index and the activity patterns overall without explicitly focusing on any one aspect of the curves. In some instances, however, there is a delay effect between the policy enactment and the response.  We might see that the slope of the stringency index suddenly increases and three days later that change is mirrored by the activity patterns.  In addressing \ref{rq3} we must first test to determine if there is a quantifiable temporal lag between government action and population response.  If a lag response exists, we can then identify which countries demonstrate the smallest or largest response lags.

The lag response analysis was done using a signal processing approach, calculating the cross-correlation as a measure of displacement similarity between each place-based activity pattern and the stringency index for each country.  The result is a vector containing the cross-correlation of the input vector based on lag.  For example, if there is no significant lag between the stringency index and an activity pattern, the highest cross-correlation would be found at lag zero and diminish as lag increases and decrease.  Should a larger number of high cross-correlation values appear in negative lag positions, this indicates that a significant change in the stringency index occurred before a significant change in the activity pattern.  Given that our units are days, the lag steps are also days.

Rather than focus purely on the single lag value with the peak cross-correlation value, we instead calculate the number of lag days with cross-correlation values above a significance alpha value of 0.5 and split them into lag days above and below zero.  We then subtract the number of remaining negative lag days from the number of positive lag days to produce our average lag for that activity pattern--stringency index pair.  This approach is intended to be more robust than looking for the single highest cross-correlation value as it considers skewness in the lag distribution.   This is a weighted approach to identifying the peak lag day as a function of all lags above the threshold avoiding an approach that ignores neighboring lag values in favor of a single peak. 

\begin{figure}[h]
	\centering
	\includegraphics[width=1\textwidth]{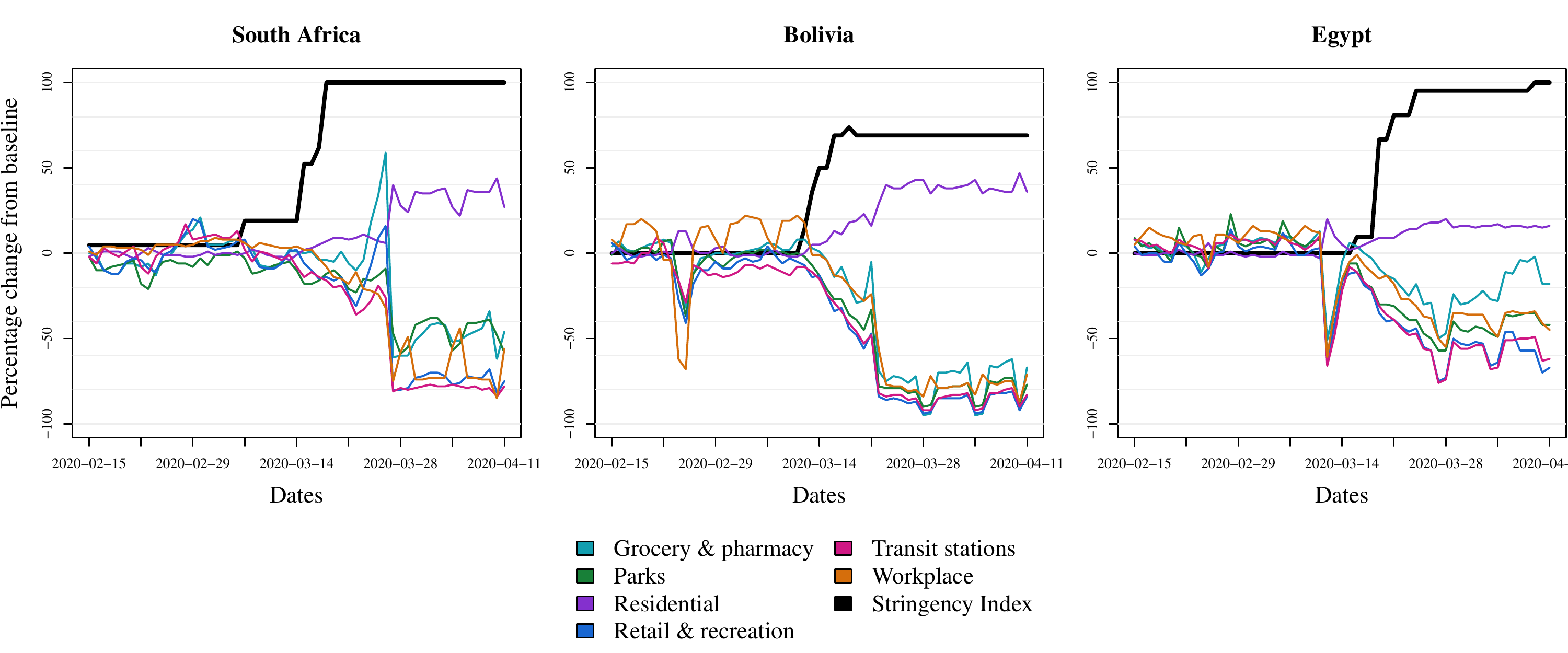}  
	\caption{Countries with high cross-correlation lag. South Africa (-8.5), Bolivia (-5.2), Egypt (+6). }
	\label{fig:toplag}
\end{figure}

Similar to our previous analysis, cross-correlation analysis was done between each activity pattern and the stringency index for each country and then averaged across all five activity patterns (excluding \textit{Parks}) producing an average, signed lag response in days.  The countries were then ranked based on the lag response.  Three countries with the largest lag---as reported by a significant change in the stringency index occurring well before or after the citizen activity response---are shown in Figure~\ref{fig:toplag}. The vast majority of countries reported a negative lag response with a mean of -2.4 days.  In a select few instances the lag response was positive. Most of these could be explained by significant non-COVID-19 related events.  For instance, Egypt's positive lag is due to a severe weather event on March 12, 2020.  Removing this event results in a lag of~-4.0.

The average lag time for each activity pattern was calculated across all countries and is shown in Table~\ref{tab:avgcc}. The high standard deviation for each of these categories speaks to the high degree of global variability in this assessment measure.  The \textit{Grocery \& pharmacy} category reports the highest average lag time including the largest standard deviation.  As many grocery stores were labeled essential services in many countries, and therefore not mandated to close, people could choose to go to a grocery store immediately, or a few days after, to stock up on supplies in response to an increase in government COVID-19 related restrictions.

\begin{table}[h]
	\centering
	\small
	\begin{tabular}{lrrr}
		\hline
			\textbf{Category}    & \textbf{Mean} & \textbf{Median} & \textbf{Standard deviation} \\
			\hline
			Retail \& recreation & -1.6          & -2              & 2.4                        \\
			Transit stations     & -1.7          & -2              & 2.6                        \\
			Parks                & -2            & -2              & 2.8                        \\
			Residential          & -2.3          & -2              & 2.4                        \\
			Workplace            & -2.6          & -3              & 2.5                        \\
			Grocery \& pharmacy  & -4.2          & -4              & 3.2                       \\ \hline
	\end{tabular}

\caption{Average lag time (days) between a significant change in the activity patterns and the stringency index, by place-based activity category.}
\label{tab:avgcc}
		\end{table}



%
%
%

\subsection{Subregional variability within countries}

Of the 108 countries with suitable place-based activity data for our analysis, 46 of them also contain sub national regions (e.g., provinces or states) with their own activity patterns.  While the stringency index is only reported at the national level, we thought it important to investigate the variation in place-based activity responses between subregions in the same country.  

To accomplish this task we extracted all subregional activity data and again split them by category.  Cosine similarity was computed for each subregion to each other subregion for each category producing six matrices per country.  The standard deviation were taken for each category matrix in each country.  Finally, we calculated the mean standard deviation across all categories in a country to produce a single ``within country'' variation value on which to compare nations.  The purpose of this analysis was to determine the degree to which countries vary in their response to government action, within their own borders.  The results speak to the power afforded national governments during crises.  A low standard deviation indicates that communities within the same country responded similarly. A high standard deviation implies that communities in different subnational regions responded differently, possibly heeding advice from local governments instead of at the national level.   

The results of this subregional similarity analysis were compared to the results of our first cosine similarity assessment approach, namely the stringency index to country level activity average similarities, using Kendall's $\tau$ method.  The resulting concordance coefficient was 0.41 (p $<$ 0.01) indicating that there is reasonably high positive correlation between the two results (\ref{rq4}).  This indicates that countries whose residents respond to national government policy actions by reducing activities are also more likely to respond consistently across subnational regions.

To this point, two countries that exemplify this difference are the \textit{United States} and \textit{New Zealand}.  The United States ranked as having one of the highest sub-national region standard deviations, which is to be expected given the authority placed on state governors during the COVID-19 pandemic and the hands-off approach of the federal government~\cite{conversation2020}.  The similarity of the United States' country-level activity response to the stringency index was in the lower 30th percentile of all countries in our dataset.  In contrast, New Zealand ranked quite high in both measures reflecting the perceived  authority of the national government and commonality of the response from residents across subregions within the country.  Overall the countries with the highest variation in subregional response were Nigeria, Uruguay, Australia, USA, and Canada.  The countries with the least variation were clustered predominantly in Europe with France, Italy, and Spain topping the list.


\subsection{Spatial relations}


Next, we investigate how our three measures (cosine similarity, lag response, subregional variation) align with the spatial relationships of the countries themselves. First, we take a high level approach by calculating the Euclidean distance between each pair of country response vectors.  Each vector is comprised of the six activity categories in each country.  Three \textit{response similarity distances} are computed for each pair of country vectors.  We then calculate the shortest geographic distance between each pair of countries in our dataset.\footnote{Using the PostGIS St\_DistanceSphere function}  Using the rank correlation approach, we computed the degree to which the geometric distance correlates with each of the response similarity distances.  Given the non-normal spatial distribution and size of countries, it is not surprising that the results, while significant, 
reported very low positive Kendall's $\tau$ values for all three measures (cosine similarity, lag time, subregion variability).

We then explore countries based on the continent to which they belong.  The average similarity distance values for each of our three measures are reported in Table~\ref{tab:continents}.  These values are split by continent and also reported overall within the same continent and between countries from different continents.  In general, there is a slight difference between the average similarity distance between countries within the same continent and those between continents.  The difference, however, is less than one might expect implying, as reported by the previous correlation values, that the spatial distribution explains little of the variability in country responses. There are, however, substantial differences between continents.  For instance, the cosine similarity distance of countries within South America is quite low indicating a high degree of similarity between countries within this continent, at least as reported by their response to government policy actions.  The number increases for lag response suggesting that though they responded similarly, there was greater variety in the lag response time than in Europe, for example.  This pattern is repeated for Africa.   The reverse, however, is true for Oceania though it must be noted that data for only three countries were used in this analysis and Papua New Guinea had a significantly different cosine similarity response than Australia and New Zealand.  These numbers should be interpreted relative to one another within the same measure, not as absolute values or compared across measures.

\begin{table}[h]
		\small
	\centering
	\begin{tabular}{lrrr}
		\hline
		{\color[HTML]{000000} \textbf{Continent}}  & {\color[HTML]{000000} \textbf{Cosine Similarity}} & {\color[HTML]{000000} \textbf{Lag Response}}  & {\color[HTML]{000000} \textbf{subregion Variability}} \\
		\hline
		{\color[HTML]{000000} Asia}                & {\color[HTML]{000000} 0.564 (0.339)}              & {\color[HTML]{000000} 0.578 (0.501)}          & {\color[HTML]{000000} 0.258 (0.230)}                    \\
		{\color[HTML]{000000} Africa}              & {\color[HTML]{000000} 0.327 (0.300)}              & {\color[HTML]{000000} 0.730 (0.645)}           & {\color[HTML]{000000} 0.485 (0.465)}                   \\
		{\color[HTML]{000000} Europe}              & {\color[HTML]{000000} 0.723 (0.628)}              & {\color[HTML]{000000} 0.412 (0.382)}          & {\color[HTML]{000000} 0.309 (0.319)}                   \\
		{\color[HTML]{000000} North America}       & {\color[HTML]{000000} 0.408 (0.325)}              & {\color[HTML]{000000} 0.596 (0.530)}           & {\color[HTML]{000000} 0.231 (0.262)}                   \\
		{\color[HTML]{000000} Oceania}             & {\color[HTML]{000000} 0.893 (0.799)}              & {\color[HTML]{000000} 0.462 (0.462)}          & {\color[HTML]{000000} 0.525 (0.525)}                   \\
		{\color[HTML]{000000} South America}       & {\color[HTML]{000000} 0.126 (0.084)}              & {\color[HTML]{000000} 0.543 (0.534)}          & {\color[HTML]{000000} 0.288 (0.291)}                   \\
		\hline
		{\color[HTML]{000000} \textit{Same Continent}}  & {\color[HTML]{000000} \textit{0.546 (0.362)}}     & {\color[HTML]{000000} \textit{0.553 (0.464)}} & {\color[HTML]{000000} \textit{0.312 (0.319)}}          \\
		{\color[HTML]{000000} \textit{Different Continent}} & {\color[HTML]{000000} \textit{0.595 (0.395)}}     & {\color[HTML]{000000} \textit{0.578 (0.513)}} & {\color[HTML]{000000} \textit{0.351 (0.333)}} \\
		\hline        
	\end{tabular}

	\caption{Mean distance values between countries within the same continents for each of our similarity measures (median in parentheses).  Overall averages for within the same continent and between different continents are shown in the last two rows. }
	\label{tab:continents}
\end{table}

\begin{figure}[h]
	\centering
	\includegraphics[width=1\textwidth]{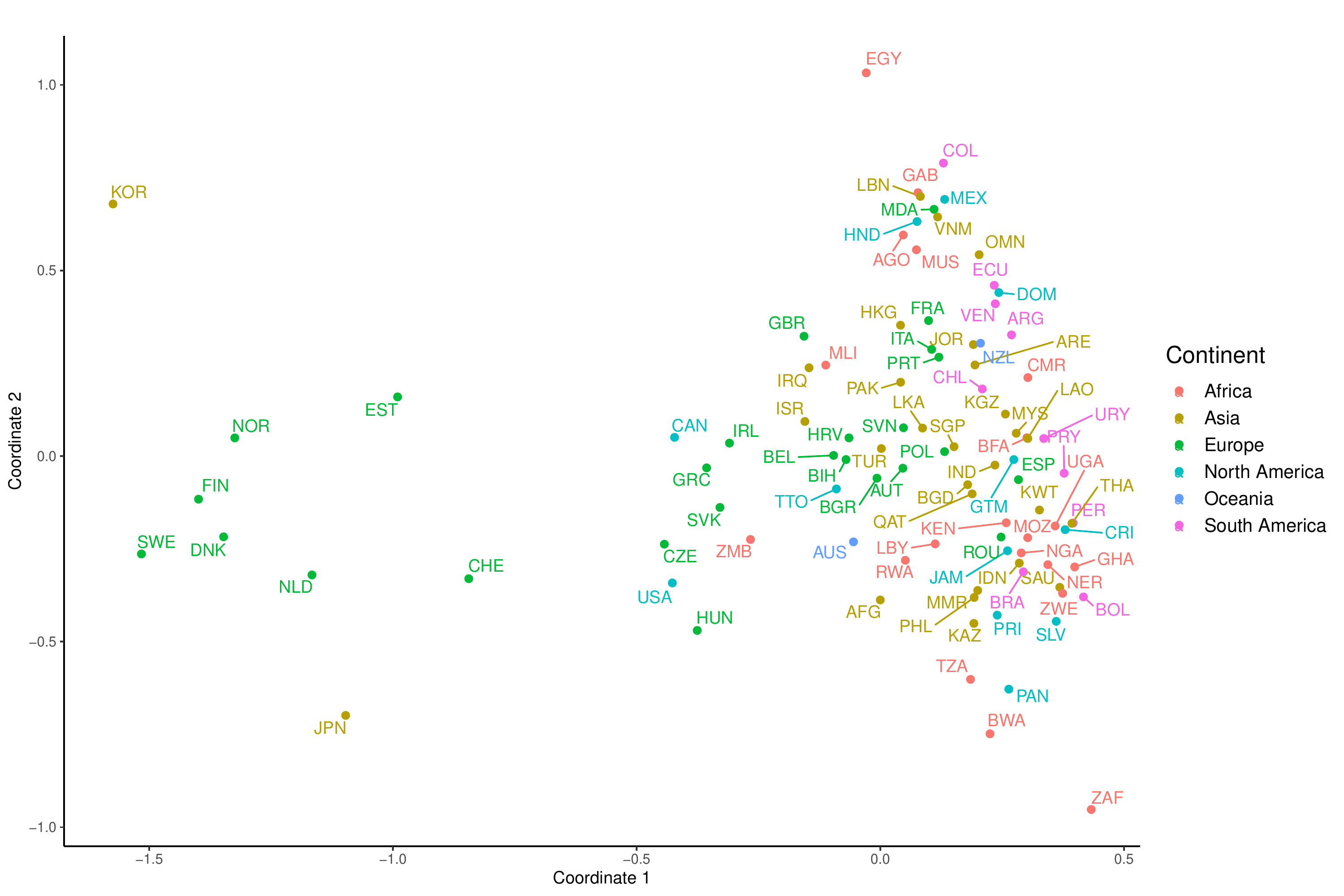}
	\caption{Multidimensional scaling of the cosine similarity and lag response time values for the six place-based activity patterns. Countries are colored by their continent in the online version of this manuscript.}
	\label{fig:mds1}
\end{figure}

Taking a multidimensional scaling approach, we compress the cosine similarity and lag response distances (again based on vector similarity) into two dimensions in order to visually represent the similarities and differences between the countries in a 2D plot.  Figure~\ref{fig:mds1} shows the results of this approach with countries assigned colors based on continent.  This figure nicely visualizes a few important findings.  The first is that there is a large cluster of countries that are all quite similar in their place-based activity response to COVID-19 policy changes.  It is difficult to single out any one specific country but we do see a mix of countries from different continents. Second, we can identify a number of countries that exist outside of the main cluster and many of these countries are from the same continent, and in some cases, the same subregion within the continent, namely Scandinavia.  Last, we can observe the outliers.  Our previous analysis indicated that South Korea and Japan responded very differently from many other countries, likely due to the fact that their governments responded before the temporal analysis window of our activity dataset.  We also see that South Africa and Egypt presented very different responses, both of which have been mentioned in previous sections.  While multidimensional scaling is a good visualization tool for displaying similarity between entities such as these, we acknowledge that it is a dimension reduction technique and so does not fully represent the nuances within the data.

\begin{figure}[H]
	\centering
	\includegraphics[width=1\textwidth]{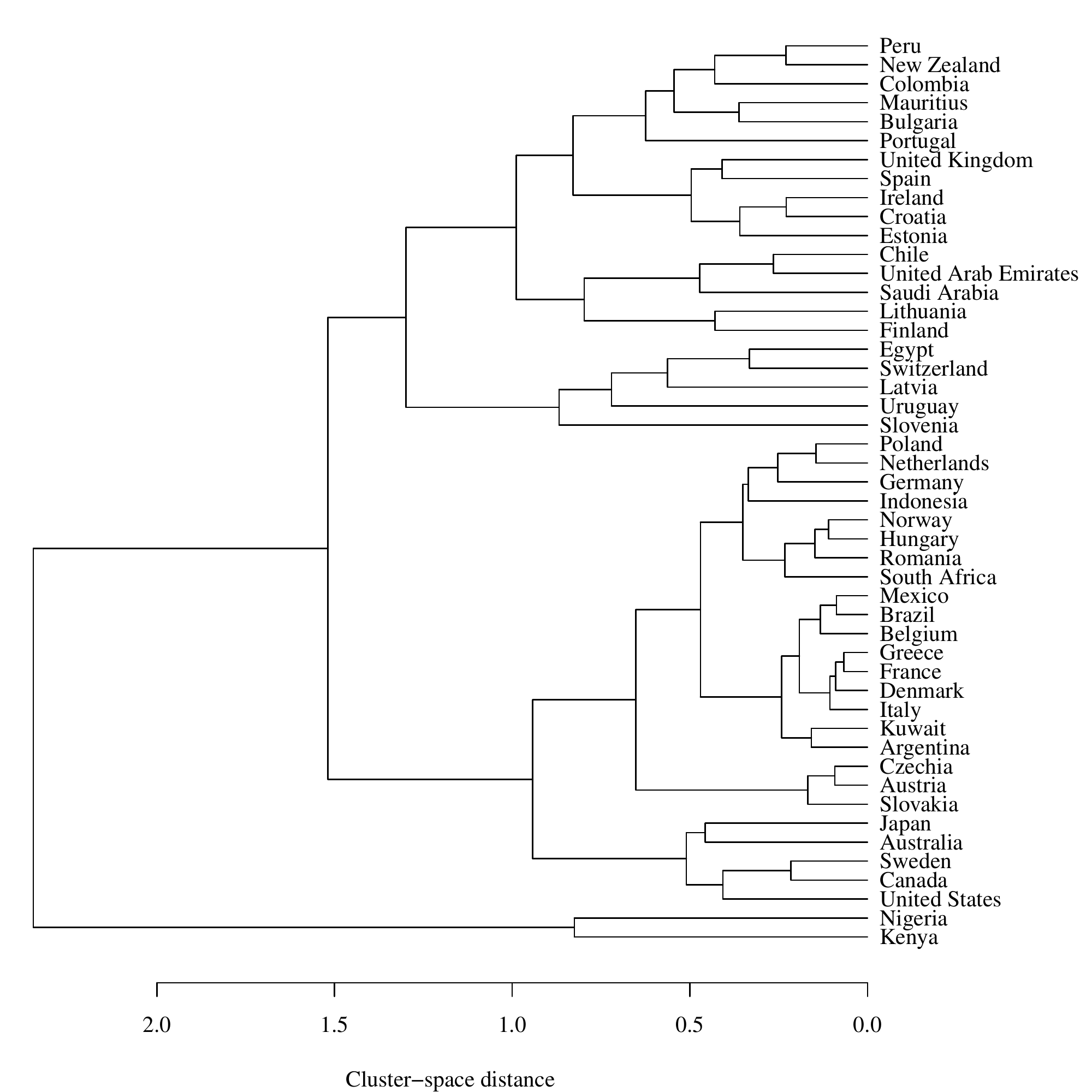}
	\caption{Dendrogram representation of the similarity between countries based on subregional variation within a country.}
	\label{fig:dendrogram}
\end{figure}

Increasing the spatial resolution of our analysis, we assess the similarities in mobility responses between countries that share a border with those that do not.  For our cosine similarity measure, those countries that shared a border reported a mean similarity distance of 0.378.  Those that do not share a border reported a mean distance of 0.589.  A similar difference in mean distance was found for the lag response and the subregion variation measures.  In pulling apart these distance values for each country pair, we found this to be the case in most countries but with a few notable exceptions.  For instance, through this approach, Germany is shown to be quite dissimilar from its neighbors based on cosine similarity, which mirrors what has been reported in the media~\cite{bennhold2020german,germany2020}.  The United States and Mexico were also quite dissimilar in their responses, considerably different than the United States and Canada.  The most similar countries that shared a border according to our cosine similarity measure are on the continent of South America, namely Columbia and Venezuela, Ecuador and Peru.  The largest lag difference in neighboring countries was also identified in South America in Columbia and Panama, as well as Central American Honduras and El Salvador. The neighboring countries that were most similar with respect to lag response time are India and Bangladesh, with Hungary and Romania also being quite close. With respect to the subregion variation analysis, the most similar neighboring countries were Italy and France (also highly similar for cosine similarity) followed by the Czechia and Slovakia.  The largest variance was also found in Europe between Italy and Slovenia as well as Switzerland and France.  Figure~\ref{fig:dendrogram} presents a visual representation of these similarities and differences through a dendrogram based on subregional variability.  Again, this is a subset of all countries where Google published subregional activity patterns.  One of the interesting things that this visualization shows is the degree to which Kenya and Nigeria (bottom two countries) are dissimilar from all other countries. This hierarchical clustering approach suggests that should all the countries be clustered into two groups based on their subregional variation, that Kenya and Nigeria would make up one cluster while all other countries would be in the second cluster.


\section{Correlations with other indices}
\label{sec:indices}

The analysis thus far has focused on the relationships between government policy action, as proxied by the stringency index, and the place-based activity response, as proxied by Google's location services-based community mobility patterns.  In this section, we explore the relationships between the findings of our analyses, country-specific attributes, and third-party global indices (\ref{rq5}).

\subsection{Country attributes, COVID-19 cases, and deaths}

One possible reason for the differences we see between countries could simply be attributed to basic information about the country such as size and population.  To further investigate this we compute Kendall's $\tau$ and find no significant rank correlation between our mean cosine similarities and population, area, or population density. 
Similarly, there is no significant rank correlation between response lag and these three variables. 
Turning to the variation within country results, we do find a significant negative correlation between population and subregional variation ($\tau$=-0.243; p$<$0.05) indicating that as the population of a country increases the variability of activity between regions decreases.  No significant correlation was found with area or population density.  

Turning our attention to the immediate discussion of COVID-19, we investigated the relationship between our computed similarity values, the number of confirmed COVID-19 cases per capita, and number of deaths due to COVID-19 per capita, by country.   These values were taken on the last date of our time window April 11, 2020.  There was a positive correlation for each of these variables with our cosine similarity measure, $\tau$=0.181 (p$<$0.01) and $\tau$=0.160 (p$<$0.05), respectively.  Similarly, there was a negative correlation with response time lag,  $\tau$=-0.177 (p$<$0.01) and $\tau$=-0.145 (p$<$0.05), respectively. Finally, variation between subregions shows no significant concordance with number of confirmed cases, but slight negative correlation ($\tau$=0.165; p$<$0.1) with number of deaths.

On this, one might reasonably ask whether the changes in people's place-based activity were due to increases in confirmed cases or deaths rather than government policy responses to the disease.  To address this we return to the stringency index and compare this index to the number of confirmed cases and deaths over time by country.  Given the difference in magnitude (stringency index is out of 100, while number of cases and deaths are limited only by the population of a country), we chose to measure the rank correlation of the variables rather than the cosine similarity.  As was reported in~\cite{oxford2020}, there is a strong positive correlation between cases, deaths, and the stringency index.  For our purpose, however, we find that the rank correlation between the average activity pattern and either confirmed cases ($\tau$=0.610) or deaths ($\tau$=0.569), by country, is lower on average than the correlation between activity patterns and the stringency index ($\tau$=0.631).  This slightly larger $\tau$ value suggests that a change in place-based activity patterns is more likely a response to government intervention than to reported numbers of cases or deaths, though the three are obviously highly correlated.

\subsection{United Nations Human Development Index}
\label{sec:unhdi}

Next we compared the results of our three different similarity measures with the \textit{United Nations Human Development Index (HDI)}.  This index consists of a large, and broad range of indicators but we restricted our analysis to the top five as identified by the UN, namely \textit{HDI rank}, \textit{Life expectancy at birth}, \textit{Expected years of schooling}, \textit{Mean years of schooling}, and \textit{Gross national income per capita}.

We found no significant correlations between the cosine similarity and any of the top HDI indicators including overall rank.  There was, however, positive correlation ($\tau$=0.156 (p $<$ 0.05)) between the HDI rank and our response lag measure suggesting that countries with a lower HDI rank demonstrated a greater lag response time to government policy action.  Three of the HDI indicators also correlated significantly with lag response time, namely \textit{Life expectancy at birth} ($\tau$=-0.191; p $<$0.01), \textit{Expected years of schooling} ($\tau$=-0.145; p$<$0.05), and \textit{Gross National Income per capita} ($\tau$=-0.126; p$<$0.1).  Comparing HDI to our standard deviation of similarity between subregions, we discovered no significant correlation with overall HDI rank or any contributing indicators. 


\subsection{World Bank World Development Indicators}

We next looked at a series of \textit{World Development Indicators (WDI)} from the \textit{World Bank}.  Of the 284 indicators that were reported by the World Bank, only five of them resulted in significant (p$<$0.1) correlation with our cosine similarity approach.  They are \textit{Financing through local equity market}, \textit{Quality of land administration}, \textit{Pupil to teacher ratio in primary education}, \textit{Diversity of workforce}, \textit{Hiring and firing practices}.  All $\tau$ values were quite low with the first three indicators showing positive correlation and the last two being negative.  

By comparison, 97 of the WDI significantly (p$<$0.1) correlated with our computed response lag with 25 of them reporting high rank correlation significance (p$<$0.01). Of these, the most interesting, and highest correlation coefficients were found in \textit{Life Expectancy} and \textit{Infant Mortality Deaths}, findings that reflect those of our comparison with the United Nations HDI indicators.  Additionally, cases of Tuberculosis and HIV/AIDS also positively correlated with lag time indicating that an increase in lag response occurred in countries with high numbers of these diseases.

Comparing WDI to our standard deviation of similarity between subregions, 53 of the WDI significantly (p$<$0.1) correlated with our measure with 10 of them reporting high rank correlation significance (p$<$0.01). The highest positive rank correlations were computed for Gross Domestic Product (GDP), and Foreign and Domestic market size indicating that the larger the variance in activity responses within a country, the larger the GDP and market sizes of the country.

\subsubsection{Transparency indices}

Finally, we looked at three different indices broadly related to the topic of transparency.  The first of these indices, the Corruption Perception Index, did not significantly correlate with any of our three computed measures.  Similarly, there were no significant relationships between the \textit{World Press Freedom Index} published by \textit{Reporters Without Borders}.  Lastly, we revisit one of the indices published by the World Bank, namely \textit{public trust in politicians}.  As was reported by omission in the World Bank section, this indicator did not significantly correlate with any of our three measures.  These findings broadly suggest that overall, trust in government and transparency did not play a substantive role in people's reaction to government policy.


\section{Discussion \& Conclusions}
\label{sec:discussion}

The purpose of the research presented in this paper is to identify the similarities and differences in how the inhabitants of different countries responded to government policies enacted during the spread of the COVID-19 pandemic.  In this section we discuss the findings as well as the limitations in our approach and future work.  We also describe an online tool for researchers and the public to further explore these data themselves.  

In addressing our first research question, \ref{rq1}, we investigated the relationship between the policy actions taken by national governments and the place-based activity response from people living the countries.  Through our analysis we found that there is indeed a quantifiable correlation between the two.  Using two different measures, we demonstrated that an increase in the COVID-19-related policies introduced by governments lead to a decrease in the mobility and activity of their population.  We further demonstrated that the correlation between these two are greater than the correlations with the underlying causes of the policies, namely confirmed COVID-19 cases, and deaths.  Investigating the countries themselves we found there to be significant variability in responses from country to country suggesting that some countries have a greater ability to actually enact envisioned policies than others. 

In comparing the six different place-type activity categories (e.g., Grocery \& pharmacy), we discovered that some types of activities were more responsive to policy changes than others (\ref{rq2}).  Transit and residential-based activities were most responsive while Grocery \& pharmacy and Parks showed the highest degree of variability between countries.  These results were supported by the findings from \ref{rq3}, namely that there is a measurable lag between policy action and activity response.  On average the lag response equated to approximately 2.4 days across countries and activity categories, but again the Grocery \& pharmacy category demonstrated the highest degree of variability between countries.  That this activity category would be the most volatile is reasonable in that groceries and pharmacies provide essential food and medication.  While many of the activities related to the other categories could be lawfully enforced through closures, establishments offering access to essential supplies remained open, thus permitting visits and reflected through increased activity.  The Workplace category, while not as volatile as the Grocery \& pharmacy category also varied a bit between countries.  In some countries we noticed weekly patterns in the data indicating that people were still going to work in some countries, even while the transit category remained low and residential was high.  This requires further investigation but points towards the variability between country responses, equity issues, and enforcement of policies. We also identified considerable differences in lag time from country to country.  Some countries, such as New Zealand reported virtually no lag between government policy and community response, whereas others, e.g., South Africa, demonstrated a substantial lag (mean 8.5 days).  Of our three similarity measures, lag response time demonstrated the highest correlation with global indices including development indicators such as life expectancy, GDP, and market size.

Further to our goal of understanding the relationship between government policy and activity response, we investigated the variability of responses between subregions within countries.  Using a subset of countries for which Google reports subnational activity patterns, we calculated the variability of responses for each subregion to the national government policies.  We answered \ref{rq4} by showing that there are quantifiable differences in the variability of activity responses within a country.  The ranking of these countries correlated with country-level cosine similarity and lag response values suggesting that countries who's residents responded quickly and with a similar degree of magnitude to the policy actions, were also more likely to respond similarly across subregions.

In addressing our final research question, \ref{rq5}, we compared our list of countries, ordered by our three assessment measures to a range of global indices developed and curated by leading multi-national organizations.  
A country's Gross Domestic Product, Life Expectancy, and Infant Mortality Deaths showed the highest correlations and suggest that there is a link between peoples' activity responses and a country's level of development. Surprisingly, none of the transparency indices correlated with our assessment measures indicating that public trust or perceived corruption in government had little impact on whether or not residents of a country followed their government's advice. 

This work is not without limitations.  The most notable limitation is the lack of transparency in how the Google Community Mobility dataset was created.  While this is a fantastic resource for researchers, very little is known about the data collection methodology, sample sizes, geolocation accuracy, etc. As researchers we are working under the assumption that the data released publicly is valid and is a representative sample of mobile device users around the world.  This is a large assumption, but one we must live with given the limited availability of COVID-19 related activity data.  Additional contextual factors may have played a role in assessing the similarity of activity responses as well.  We know weather, for instance, affects mobility and place-based activity.  As the COVID-19 pandemic spread, northern hemisphere countries were just beginning to enter Spring which itself causes a change in activity behavior.  Parks are likely the category most impacted by this, but other activity categories may have been impacted. Similarly, the partisan politics surrounding government policies and activity responses were largely ignored in this analysis.  A wide range of political hierarchies exist with many governments wielding most of the control at the national level and others delegating health-related policy to local districts and municipalities.  Indices that report on the political underpinnings of a country would be useful to include in future work. 






\subsection{Future work}

As we are currently in the midst of this pandemic, there are a lot of unknowns and it is not yet clear what the outcomes of COVID-19-related government policies, nor activity reductions will be.  Future work on this topic will extend to analyzing people's response to relaxation of restrictive government policies and re-opening of economies.  It would be interesting to determine if residents respond to policy relaxation in similar ways. Additional datasets are being made available everyday and many of these datasets could be used in future analysis.  While we compared activity response to three different measures, namely stringency index, confirmed cases, and deaths, more government-related response indices would be worth investigation should they become available.

\subsection{Interactive visual analytics platform}

\begin{figure}[h]
	\centering
	\includegraphics[width=0.9\textwidth]{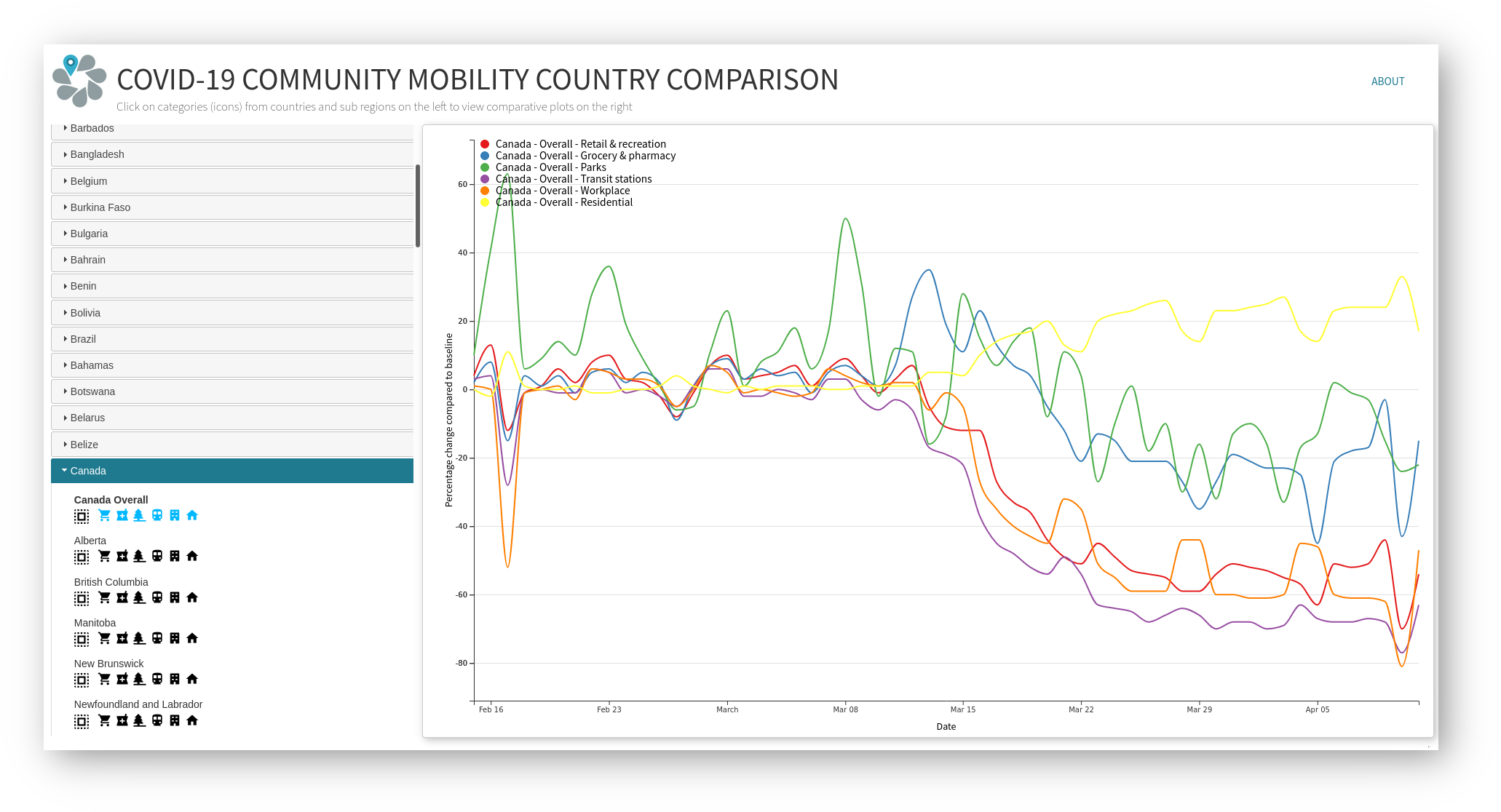}
	\caption{Interactive COVID-19 community mobility country comparison platform.}
	\label{fig:webportal}
\end{figure}

During our analyses, we developed a visual analytics platform for quickly comparing country responses across the range of activity categories (Figure~\ref{fig:webportal}).  We decided to share this platform publicly with the goal of allowing the public and researchers to explore these data and relationships.  

The platform ingests all place-based activity data as reported by Google, at the country level (and subnational level when available).  Users are encouraged to toggle between the different categories and regions to compare activity responses. The platform was built using JavaScript and the D3 framework\footnote{\url{https://d3js.org/}} and is freely available to explore at \url{https://platial.science/covid19}.

\subsection{Conclusions}

The COVID-19 pandemic will undoubtedly have a lasting impact on the global economy and society at large.  While we are racing to coming to terms with the global impact of this crisis, it is important to investigate the early patterns emerging from the spread of this pandemic in order to help us plan for future crises or a potential second wave of this current disease. Our focus in this paper was on gaining a better understanding of the relationship between the actions taken by national governments, with respect to COVID-19-related policy, and the response of their inhabitants, specifically as it relates to their activities.  We showed considerable differences exist between countries.  Three approaches were used to quantify the relationship between policy and response. Though countries varied between measure, commonalities emerged.  A number of the similarities and differences between countries can be explained through correlations with global indices such as market size, or life expectancy.  Not surprisingly, countries that share a border were also more similar in their activity responses.  As this pandemic continues to spread and we look towards an uncertain future, understanding how people respond to their government is an important step in combating this crisis.





\bibliographystyle{elsarticle-num}
\bibliography{main.bib}








\end{document}